\documentclass{article}
\usepackage{amssymb}
\usepackage{amsfonts}
\usepackage{amsmath}

\begin{document}

\title{The geometry of noncommutative space-time}
\author{R. Vilela Mendes \\
Centro de Matem\'{a}tica e Aplica\c{c}\~{o}es Fundamentais, University of
Lisbon\\
C6 - Campo Grande 1749-016 Lisboa, Portugal}
\date{ }
\maketitle

\begin{abstract}
Stabilization, by deformation, of the Poincar\'{e}-Heisenberg algebra
requires both the introduction of a fundamental lentgh and the
noncommutativity of translations which is associated to the gravitational
field. The noncommutative geometry structure that follows from the deformed
algebra is studied both for the non-commutative tangent space and the full
space with gravity. The contact points of this approach with the work of
David Finkelstein are emphasized.
\end{abstract}

\section{Introduction}

I first met David Finkelstein when, as a graduate student at Austin, went to
a summer school where David was one of the lecturers. Further to his
excellent lectures, I was deeply impressed by his warm readiness to meet and
answer the questions of the students. Much later, through a common friend,
Eric Carlen, I was introduced to his simplicity approach to physical
theories and conversely he became aware of my approach to noncommutative
space-time through deformation theory, to which he then gave generous
reference in his papers.

We had planned to meet when he once passed by Lisbon, but at the time I was
in France and unfortunately missed that chance. Nevertheless the
correspondence we exchanged and the reading of his papers have been a
constant source of inspiration for the exploration of uncharted and
sometimes unpopular territory.

In this paper, which I dedicate to the memory of David Finkelstein, the
focus will be on the geometry of the noncommutative space time that follows
from the (stable) deformed Poincar\'{e}-Heisenberg algebra. The many contact
points with Finkelstein approach to these problems will be put into
evidence. Among his many important contributions in different fields, our
contact point came about in the study of modifications of the space-time
algebra, which Finkelstein approached through the requirement of simplicity
of the algebras \cite{Fink6} \cite{Segal}, whereas I have used a
deformation-stability principle. Of course, in addition to stability, there
are other arguments in favor of simple algebras, in particular the spectrum
of its representations \cite{Fink5}. For quantum physics the deformation
stability approach does indeed coincide with the simplicity approach, but
deformation-stability may well go beyond the Lie algebra realm \cite{Vilela0}%
.

\section{The stable Poincar\'{e}-Heisenberg algebra}

The Poincar\'{e}-Heisenberg algebra is deformed \cite{Vilela1} to the stable
algebra $\Re _{\ell ,R}=\{M^{\mu \nu },p^{\mu },x^{\mu },\Im \}$ defined by
the commutators%
\begin{equation}
\begin{array}{lll}
\lbrack M^{\mu \nu },M^{\rho \sigma }] & = & i(M^{\mu \sigma }\eta ^{\nu
\rho }+M^{\nu \rho }\eta ^{\mu \sigma }-M^{\nu \sigma }\eta ^{\mu \rho
}-M^{\mu \rho }\eta ^{\nu \sigma }) \\ 
\lbrack M^{\mu \nu },p^{\lambda }] & = & i(p^{\mu }\eta ^{\nu \lambda
}-p^{\nu }\eta ^{\mu \lambda }) \\ 
\lbrack M^{\mu \nu },x^{\lambda }] & = & i(x^{\mu }\eta ^{\nu \lambda
}-x^{\nu }\eta ^{\mu \lambda }) \\ 
\lbrack p^{\mu },x^{\nu }] & = & i\eta ^{\mu \nu }\Im \\ 
\lbrack x^{\mu },x^{\nu }] & = & -i\epsilon _{4}\ell ^{2}M^{\mu \nu } \\ 
\lbrack p^{\mu },p^{\nu }] & = & -i\frac{\epsilon _{5}}{R^{2}}M^{\mu \nu }
\\ 
\lbrack x^{\mu },\Im ] & = & i\epsilon _{4}\ell ^{2}p^{\mu } \\ 
\lbrack p^{\mu },\Im ] & = & -i\frac{\epsilon _{5}}{R^{2}}x^{\mu } \\ 
\lbrack M^{\mu \nu },\Im ] & = & 0%
\end{array}
\label{A1}
\end{equation}%
which some authors now call the \textit{stable Poincar\'{e}-Heisenberg
algebra}

The stable algebra $\Re _{\ell ,R}$, to which the Poincar\'{e}-Heisenberg
algebra has been deformed, is the algebra of the $6-$dimensional
pseudo-orthogonal group with metric 
\begin{equation}
\eta _{aa}=(1,-1,-1,-1,\epsilon _{4},\epsilon _{5}),\bigskip \ \epsilon
_{4},\epsilon _{5}=\pm 1  \label{A2}
\end{equation}%
with the identifications%
\begin{equation}
\begin{array}{ccc}
p^{\mu } & = & \frac{1}{R}M^{\mu 4} \\ 
x^{\mu } & = & \ell M^{\mu 5} \\ 
\Im & = & \frac{\ell }{R}M^{45}%
\end{array}
\label{A3}
\end{equation}%
Both $\ell $ and $R$ have dimensions of length. However they might have
different physical status and interpretation. Whereas $\ell $ might be
considered as a fundamental length and a constant of Nature, $R$, being
associated to the non-commutativity of the generators of translation in the
Poincar\'{e} group, seems to be associated to the local curvature of the
space-time manifold and therefore is a dynamical quantity associated to the
local intensity of the gravitational field.

In the tangent space one may take the limit $R\rightarrow \infty $ obtaining%
\begin{equation}
\left. \lbrack p^{\mu },p^{\nu }]\right\vert _{R\rightarrow \infty
}\rightarrow 0\hspace{2cm}\text{and\hspace{2cm}}\left. [x^{\mu },\Im
]\right\vert _{R\rightarrow \infty }\rightarrow 0  \label{A4}
\end{equation}%
all the other commutators being the same as in (\ref{A1}), leading to the
tangent space algebra $\Re _{\ell ,\infty }=\left\{ x^{\mu },M^{\mu \nu },%
\overline{p}^{\mu },\overline{\Im }\right\} \footnote{$\overline{p}^{\mu },%
\overline{\Im }$ denote the tangent space ($R\rightarrow \infty $) limits of
the operators, not be confused with the physical $p^{\mu },\Im $ operators.
According to the deformation-stability principle they are stable physical
operators only when $R$ is finite, that is, when gravity is turned on.}$\
whose consequences have been studied in a number of publications \cite%
{Vilela2}-\cite{Vilela8}. In this limit the operators $\left\{ \overline{p}%
^{\mu },\overline{\Im }\right\} $ are an Abelian set of derivations of the $%
\Re _{\ell ,\infty }$ algebra.

Finkelstein, in line with his suggestion that Clifford algebra is the
natural language for quantum physics \cite{Fink1} \cite{Fink2}, identifies
the world line of a spin $\frac{1}{2}$ particle with $N$ Clifford cells, the
usual Dirac spin being the growing tip of the world line \cite{Fink3} \cite%
{Fink4}. The space-time operators are then represented as sums of
second-order elements in the spinor $6N$ space, for example%
\begin{eqnarray}
\overset{\smile }{x}^{\mu } &=&-\chi \sum_{n=1}^{N-1}\gamma ^{\mu 4}\left(
n\right)  \notag \\
\overset{\smile }{p}^{\mu } &=&\phi \sum_{n=1}^{N-1}\gamma ^{\nu 5}\left(
n\right)  \label{A5}
\end{eqnarray}%
Then, essentially the same commutation relations as in (\ref{A1}) are
obtained, the $\epsilon _{4},\epsilon _{5}=\pm 1$ metric choice being
related to the real or imaginary nature of the simplifier (deformation)
parameters $\chi $ and $\phi $. The independent parameters are also two in
number, $N$ being constrained by%
\begin{equation}
\chi \phi \left( N-1\right) =\frac{\hbar }{2}  \label{A6}
\end{equation}

In the setting of the stable $\Re _{\ell ,R}$ algebra all variables are
represented as operators with equal footing, the space-time coordinates
themselves not having a special distinguished role. In particular the
absence of nontrivial characters, implies that space-time has no points.
Rather, the physical processes will be operations on a representation space
(a module) over the algebra. The most economic way to construct a module
would be to use free powers of the algebra itself, leading to the notion of
physical processes as operations on free modules over the algebra. The view
of physics as a process and an unifying status for the physical variables as
operators was in several forms and places vigorously proposed by David
Finkelstein.

In the following I will deal with the space-time geometry that follows from
the stable Heisenberg-Poincar\'{e} algebra $\Re _{\ell ,R}$ and its tangent
space limit $\Re _{\ell ,\infty }$.

\section{The space-time geometry and some consequences}

In the classical (commutative) case the space-time coordinates $\left\{
x^{\mu }\right\} $ are a commuting set whereas, in the deformed setting of (%
\ref{A1}), the space-time algebra becomes%
\begin{equation}
\begin{array}{lll}
\lbrack M^{\mu \nu },M^{\rho \sigma }] & = & i(M^{\mu \sigma }\eta ^{\nu
\rho }+M^{\nu \rho }\eta ^{\mu \sigma }-M^{\nu \sigma }\eta ^{\mu \rho
}-M^{\mu \rho }\eta ^{\nu \sigma }) \\ 
\lbrack M^{\mu \nu },X^{\lambda }] & = & i(X^{\mu }\eta ^{\nu \lambda
}-X^{\nu }\eta ^{\mu \lambda }) \\ 
\lbrack X^{\mu },X^{\nu }] & = & -i\epsilon _{4}M^{\mu \nu }%
\end{array}
\label{B1}
\end{equation}%
a non-commutative algebra, with $X^{\mu }=\frac{x^{\mu }}{\ell }$.

There is a clear relation between algebraic and geometric structures.
Indeed, the way one explores a space $S$ is by computing functions on it and
functions on $S$ form algebras. In the classical (commutative) case the
Gelfand-Naimark theorem states that a $C^{\ast }-$algebra $A$ is $\ast -$%
isomorphic to an algebra of functions $C_{0}\left( S_{A}\right) $ on its
Gelfand spectrum $S_{A}$ and the Serre-Swan theorem that continuous sections
of a finite dimensional vector bundle $E\rightarrow M$ are finitely
generated projective modules over $C\left( M\right) $ and every such module
is a space of sections of a vector bundle over $M$. These and other
correspondences were extended to non-commutative algebras, providing a
framework for non-commutative geometry \cite{DuBois} \cite{Connes2} \cite%
{Connes-Marcolli}.

Given an algebra $A$, the standard way to obtain the correspondent geometry
and in particular the differential algebra structure is by forming a triple $%
\left( H,\pi (A),D\right) $, where $\pi (A)$ is a representation of the
algebra in the Hilbert space $H$ and $D$ is a Dirac operator. When a
sufficient number of algebra derivations are available, the noncommutative
generalization of the geometrical notions is a natural extension of the
commutative case \cite{DuBois}. However, in general it might not be possible
to use derivations to construct by duality the differential forms because
many algebras have no derivations at all. The commutator with the Dirac
operator is then used to generate the one-forms, the Dirac operator also
providing the metric structure. Notice, however that, although the language
of spectral geometry through the $\left( H,\pi (A),D\right) $ triple may be
used as a guide, the assumptions of compactness and positive definite metric
used in most rigorous constructions do not apply to the algebras studied
here.

Depending on the sign of $\epsilon _{5}$ the algebra in (\ref{B1}) is the
algebra of $SO\left( 3,2\right) $, $\epsilon _{4}=+1$, or $SO\left(
4,1\right) $, $\epsilon _{4}=-1$. The simplest representation of these
algebras would be as differential operators in a five-dimensional Euclidean
space with coordinates $(\xi ^{1},\xi ^{2},\xi ^{3},\xi ^{0},\xi ^{4})$%
\begin{equation}
\begin{array}{rll}
M^{\mu \nu } & = & i(\xi ^{\mu }\frac{\partial }{\partial \xi _{\nu }}-\xi
^{\nu }\frac{\partial }{\partial \xi _{\mu }}) \\ 
x^{\mu } & = & \xi ^{\mu }+i\ell (\xi ^{\mu }\frac{\partial }{\partial \xi
^{4}}-\epsilon _{4}\xi ^{4}\frac{\partial }{\partial \xi _{\mu }})%
\end{array}
\label{B2}
\end{equation}%
In this setting the operators $\overline{p}^{\mu },\overline{\Im }$ have a
representation%
\begin{eqnarray*}
\overline{p}^{\mu } &=&i\frac{\partial }{\partial \xi _{\mu }} \\
\overline{\Im } &=&1+i\ell \frac{\partial }{\partial \xi ^{4}}
\end{eqnarray*}%
generating, together with $M^{\mu \nu }$ and $x^{\mu }$ the algebras of the
inhomogeneous $ISO\left( 3,2\right) $ or $ISO\left( 4,1\right) $. The
minimal Abelian set of derivations generalizing those of the commutative
case are associated to the operators $\overline{p}^{\mu },\overline{\Im }$.
Also in the commutative case, the derivations, used to construct the
differential calculus are not inner derivations, as has been proposed in
some versions of the matrix geometries \cite{DuBois}, but operations of the
Heisenberg algebra. The following maximal Abelian set $V=\{\overline{%
\partial }^{\mu },\overline{\partial }^{4}\}$ of derivations in the
enveloping algebras of the inhomogeneous groups are used%
\begin{equation}
\begin{array}{lll}
\overline{\partial }^{\mu }(x^{\nu }) & = & \eta ^{\mu \nu }\Im \\ 
\overline{\partial }^{4}(x^{\mu }) & = & -\epsilon _{4}\ell p^{\mu }\Im \\ 
\overline{\partial }^{\sigma }(M^{\mu \nu }) & = & \eta ^{\sigma \mu }p^{\nu
}-\eta ^{\sigma \nu }p^{\mu } \\ 
\overline{\partial }^{4}(M^{\mu \nu }) & = & 0%
\end{array}
\label{B3}
\end{equation}%
From this, by duality, the differential calculus is constructed \cite%
{Vilela3}. Notice that although an extra dimension is used in the
representation space, the space-time coordinates are still only four,
noncommutative ones. However the derivations in $V$ introduce, by duality,
an additional degree of freedom in the exterior algebra. The Dirac operator
is%
\begin{equation}
\overline{D}=i\gamma ^{a}\overline{\partial }_{a}  \label{B4}
\end{equation}%
with $\overline{\partial }_{a}=\left( \overline{\partial }_{\mu },\overline{%
\partial }_{4}\right) $, the $\gamma $'s being a basis for the Clifford
algebras $C\left( 3,2\right) $ or $C\left( 4,1\right) $%
\begin{equation}
\begin{array}{ccc}
\left( \gamma ^{0},\gamma ^{1},\gamma ^{2},\gamma ^{3},\gamma ^{4}=\gamma
^{5}\right) &  & \epsilon _{4}=+1 \\ 
\left( \gamma ^{0},\gamma ^{1},\gamma ^{2},\gamma ^{3},\gamma ^{4}=i\gamma
^{5}\right) &  & \epsilon _{4}=-1%
\end{array}
\label{B5}
\end{equation}%
This was the approach followed in \cite{Vilela3} and \cite{Vilela9}. The
representation (\ref{B2}) is an efficient tool for calculations, however it
is not irreducible.

In the commutative case, the points of the geometry are the characters, the
continuous algebra morphisms from the algebra to the complex numbers.
Noncommutative algebras have no such characters and the most elementary
geometric sets are the irreducible representations. Hence, the elementary
space-time structures compatible with the (tangent space) deformed algebra
are to be obtained from the irreducible representations of the groups $%
SO\left( 3,2\right) $ or $SO\left( 4,1\right) $. For future reference, those
of $SO\left( 3,2\right) $ are listed in the Appendix.

Deformation stability, or Lie algebra simplicity, as strongly proposed by
Finkelstein, leads to modifications of the space-time algebra of the type
described before, in particular the noncommutativity of the coordinates and
a radical modification of the view of physical processes as happening in the
background of a smooth space-time manifold. Experimental observation of the
effects of these modifications will of course depend on the size of the
deformation parameter $\ell $. Two types of effects were predicted: those
that depend simply on the noncommutativity of the variables and those that
depend on the dimension of the differential algebra. Effects of the first
type are associated for example to modifications of the phase-space volume 
\cite{Vilela4} \cite{Vilela6} \cite{Vilela8} or to the measurement of the
velocity of wave packets, because time and space being noncommuting
operators their ratio can only be taken in the sense of expectation values 
\cite{Vilela7}.

The structure of the differential algebra and the associated Dirac operator (%
\ref{B4}) also implies the existence of two solutions for the "massless"
Dirac equation , one massless and the other of very large mass (of order $%
1/\ell $) with the same quantum numbers \cite{Vilela9}. Mixing of these
solutions might, by the seesaw mechanism, endow neutrinos with small masses.
Here it must be pointed out that Galiautdinov and Finkelstein \cite{Fink3},
following a slightly different approach, also studied modifications to the
Dirac equation. I do not feel comfortable with their interpretation of the
relation to the Higgs mass, but the fact remains that they also pointed out
the existence of large mass solutions.

Less explored is the fact that if the differential algebra has an additional
dimension then, quantum fields that are connections should have an
additional component\footnote{%
Notice that additional components are not necessarily required for spinors
because the Clifford algebras $C\left( 3,2\right) $ or $C\left( 4,1\right) $
both have four-dimensional representations.} \cite{Vilela3}.

\section{Gravity as a quantum effect}

I borrow the title of this section from the title of a preprint of David
Finkelstein. The preprint itself was unpublished, I think, but the main
ideas were published in \cite{Fink7}. According to Finkelstein "\textit{the
non-commutativity of parallel transport is a classical vestige of the
quantum non-commutativity of event momentum-energy variables}".

Indeed, when the full stable $\Re _{\ell ,R}$ algebra is considered, the
generators of translations no longer commute,%
\begin{equation}
\begin{array}{lll}
\lbrack p^{\mu },p^{\nu }] & = & -i\frac{\epsilon _{5}}{R^{2}}M^{\mu \nu }%
\end{array}
\label{C1}
\end{equation}%
Redefining $\frac{\epsilon _{5}}{R^{2}}$ as a new, gravity related,
space-time dependent field $\phi $%
\begin{equation}
\frac{\epsilon _{5}}{R^{2}}\circeq \phi  \label{C2}
\end{equation}%
\begin{equation}
\begin{array}{lll}
\lbrack p^{\mu },p^{\nu }] & = & -i\phi M^{\mu \nu } \\ 
\lbrack p^{\mu },\Im ] & = & -i\phi x^{\mu }%
\end{array}
\label{C3}
\end{equation}%
The algebra will be the same as before if $\phi $ commutes with all the
generators, that is, if it is a function of the invariants. The invariants
are those of the algebra of the $6-$dimensional pseudo-orthogonal group with
the metric in (\ref{A2}) and the identifications in (\ref{A3}). Then with
indices $a,b,\cdots \in \left\{ 0,1,2,3,4,5\right\} $ the invariants are 
\begin{eqnarray}
C_{1} &=&\sum M_{ab}M^{ab}  \notag \\
C_{2} &=&\sum \epsilon _{abcdef}M^{ab}M^{cd}M^{ef}  \notag \\
C_{3} &=&\sum M_{ab}M^{bc}M_{cd}M^{da}  \label{C4}
\end{eqnarray}%
with summation over repeated indices and $\epsilon _{012345}=+1$.

In this view, the scalar (operator) field $\phi $ appears, rather than the
metric, as the primary gravitational field. Commutativity of $\phi $ with
all the generators is an expression of the conformal covariance of
gravity-related tensors. Notice however that this is not DeSitter or
anti-DeSitter geometry. It would be if $\ell =0$ and $\phi =$constant, but
here the coordinates are noncommuting operators ($\ell \neq 0$) and $\phi $
is also an operator-valued function of the invariants $C_{i}$.

With (\ref{C3}) and (\ref{A1}) a non-commutative geometry framework for $\Re
_{\ell ,R}$ may be developed along the same lines as done before \cite%
{Vilela3} for the tangent-space $\Re _{\ell ,\infty }$ algebra.

The basic spaces to be used are the enveloping algebra $U_{\Re }$ generated
by $x^{\mu },M^{\mu \nu },p^{\mu },\Im $ plus a unit and $\Im ^{-1}$%
\begin{equation}
U_{\Re }=\left\{ x^{\mu },M^{\mu \nu },p^{\mu },\Im ,\Im ^{-1},\boldsymbol{1}%
\right\}  \label{C5}
\end{equation}%
and free modules generated by $U_{\Re }$. Because of (\ref{C3}), that is,
because the set $\left\{ p^{\mu },\Im \right\} $ is not closed under
commutation, one should take into account the full space $\mathcal{V}$ of
inner derivations corresponding to $\left\{ x^{\mu },M^{\mu \nu },p^{\mu
},\Im \right\} $, namely%
\begin{eqnarray}
\partial ^{\mu } &\longleftrightarrow &\frac{1}{i}p^{\mu }  \notag \\
\partial ^{4} &\longleftrightarrow &\frac{1}{i\ell }\Im  \notag \\
\partial ^{\mu \nu } &\longleftrightarrow &\frac{1}{i}M^{\mu \nu }  \notag \\
\partial ^{x_{\mu }} &\longleftrightarrow &\frac{1}{i}x^{\mu }  \label{C6}
\end{eqnarray}%
the correspondence symbol $\longleftrightarrow $ in (\ref{C6}) meaning that
the derivations $\partial $ act on $U_{\Re }$ in the same way as the
commutators with the operators on the right hand side.

Then the graded differential algebra $\Omega \left( U_{\Re }\right) $ is the
complex of multilinear antisymmetric mappings from the space $\mathcal{V}$
of derivations to $U_{\Re }$. $\Omega ^{0}\left( U_{\Re }\right) $ is
identified with $U_{\Re }$. Defining a basis%
\begin{equation}
\theta _{a}\left( \partial ^{b}\right) =\delta _{a}^{b}  \label{C7}
\end{equation}%
with $a,b\in \left\{ \mu ,4,\mu \nu ,x_{\mu }\right\} $, and an exterior
derivative in $\Omega \left( U_{\Re }\right) $%
\begin{eqnarray}
d\omega \left( \partial ^{a_{0}},\cdots ,\partial ^{a_{k}}\right)
&=&\sum_{i}\left( -1\right) ^{i}\partial ^{a_{i}}\left( \omega \left(
\partial ^{a_{0}}\cdots ,\widehat{\partial ^{a_{i}}},\cdots ,\partial
^{a_{k}}\right) \right)  \notag \\
&&+\sum_{i<j}\left( -1\right) ^{i+j}\omega \left( \left[ \partial
^{a_{i}},\partial ^{a_{j}}\right] ,\partial ^{a_{0}}\cdots ,\widehat{%
\partial ^{a_{i}}},\cdots ,\widehat{\partial ^{a_{j}}},\cdots ,\partial
^{a_{k}}\right)  \notag \\
&&  \label{C8}
\end{eqnarray}%
the differential of physical operators may be computed. For example:%
\begin{eqnarray}
dx^{\mu } &=&\eta ^{\mu \nu }\Im \theta _{\nu }-\epsilon _{4}\ell p^{\mu
}\theta _{4}+\left( \eta ^{\beta \mu }x^{\alpha }-\eta ^{\alpha \mu
}x^{\beta }\right) \theta _{\alpha \beta }-\epsilon _{4}\ell ^{2}M^{\alpha
\mu }\theta _{x_{\alpha }}  \notag \\
dp^{\mu } &=&-\phi M^{\nu \mu }\theta _{\nu }+\frac{\phi }{\ell }x^{\mu
}\theta _{4}+\left( \eta ^{\beta \mu }p^{\alpha }-\eta ^{\alpha \mu
}p^{\beta }\right) \theta _{\alpha \beta }-\eta ^{\alpha \mu }\Im \theta
_{x_{\alpha }}  \label{C9}
\end{eqnarray}

One also defines a contraction $i_{\partial }$ as a mapping from $\Omega
^{p}\left( U_{\Re }\right) $ to $\Omega ^{p-1}\left( U_{\Re }\right) $%
\begin{equation}
i_{\partial }\omega \left( \partial ^{a_{1}},\cdots ,\partial
^{a_{p-1}}\right) =\omega \left( \partial ,\partial ^{a_{1}},\cdots
,\partial ^{a_{p-1}}\right)  \label{C10}
\end{equation}%
and a Lie derivative%
\begin{equation}
L_{\partial }=di_{\partial }+i_{\partial }d  \label{C11}
\end{equation}

Let now $E$ be a $U_{\Re }-$left module generated by the identity $%
\boldsymbol{1}$%
\begin{equation}
E=\left\{ a\boldsymbol{1};a\in U_{\Re }\right\}  \label{C12}
\end{equation}%
From this, other modules may be obtained by projection, a projection $\Pi $
being a matrix with entries in $U_{\Re }$%
\begin{equation}
E_{\Pi }=\left\{ \psi \in E:\Pi \psi =\psi \right\}  \label{C13}
\end{equation}%
with $\sum_{i=1}^{n}\psi _{i}\Pi _{ji}=\psi _{j}$. The modules $E_{\Pi }$
are, in this non-commutative context, the $n-$dimensional quantum fields and%
\begin{equation}
\Pi \psi -\psi =0  \label{C14}
\end{equation}%
the field equations.

In $E$ one defines connections $\nabla $ as mappings $\nabla :E\rightarrow
\Omega ^{1}\left( U_{\Re }\right) \otimes E$ such that%
\begin{equation}
\nabla \left( a\chi \right) =a\nabla \left( \chi \right) +da\chi  \label{C15}
\end{equation}%
with $a\in U_{\Re }$ and $\chi \in E$. Because of (\ref{C15}), if one knows
how the connection acts on the algebra unit $\boldsymbol{1}$ one has the
complete action. Let%
\begin{equation}
\nabla \left( \boldsymbol{1}\right) =A^{i}\theta _{i}  \label{C16}
\end{equation}%
Then%
\begin{equation}
\nabla \left( x^{\mu }\right) =\nabla \left( \boldsymbol{1}x^{\mu }\right)
=A^{i}\theta _{i}x^{\mu }+dx^{\mu }  \label{C17}
\end{equation}%
The covariant derivative along $\partial $ is $\nabla _{\partial }=\left(
\nabla ,\partial \right) $ and the curvature is obtained by the commutator
of two covariant derivatives. Using (\ref{C17}) one obtains%
\begin{equation}
\left[ \nabla _{\partial ^{\alpha }},\nabla _{\partial ^{\beta }}\right]
\left( x^{\mu }\right) =\left\{ d_{\alpha }A^{\beta }-d_{\beta }A^{\alpha }-%
\left[ A^{\alpha },A^{\beta }\right] \right\} x^{\mu }+\phi \left( \eta
^{\sigma \alpha }\eta ^{\mu \beta }-\eta ^{\sigma \beta }\eta ^{\mu \alpha
}\right) x_{\sigma }  \label{C18}
\end{equation}%
with $d_{\alpha }A^{\beta }=\left( dA^{\beta },\partial ^{\alpha }\right) $.
One sees that, in addition to the curvature of the (gauge) field $A$, there
is a gravitational induced curvature associated to the stable
Heisenberg-Poincar\'{e} algebra $\Re _{\ell ,R}$, with curvature tensor,%
\begin{equation}
R^{\sigma \alpha \mu \beta }=\phi \left( \eta ^{\sigma \alpha }\eta ^{\mu
\beta }-\eta ^{\sigma \beta }\eta ^{\mu \alpha }\right)  \label{C19}
\end{equation}%
the deformation field $\phi $ appearing as an operator-valued scalar
curvature. Notice that, through (\ref{C4}), $\phi $ may depend not only on
the coordinate operators but also on the momentum and angular momentum
operators.

Here, because the algebra $\Re _{\ell ,R}$ has a sufficiently rich set of
derivations, the construction of the graded differential algebra was based
on the derivations. However there are other standard ways to construct the
differential algebra which do not rely on the existence of derivations. One
of them uses an operator $D$ and defines $p-$forms as%
\begin{equation}
\omega =\sum a_{0}\left[ D,a_{1}\right] \cdots \left[ D,a_{p}\right] 
\label{C20}
\end{equation}%
Let $\Gamma $ be the following set of 15 four-dimensional gamma matrices%
\begin{equation}
\Gamma =\left\{ \gamma ^{\mu },\left( i\right) ^{\frac{1-\epsilon _{4}}{2}%
}\gamma ^{5},\gamma ^{\mu \nu }=\frac{1}{2}\left[ \gamma ^{\mu },\gamma
^{\nu }\right] ,\gamma ^{\mu }\gamma ^{5}\right\}   \label{C21}
\end{equation}%
and the operator%
\begin{equation}
D=i\Gamma _{a}\partial ^{a}  \label{C22}
\end{equation}%
where $\left\{ \partial ^{a}\right\} $ is the set $\Gamma $ of 15
derivations, listed in (\ref{C6}). Then from (\ref{C20}) and (\ref{C22}) one
sees that the differential algebra structure obtained by (\ref{C20}) is
identical to the one obtained from the derivations. The operator $D$ may be
identified as the Dirac operator of the theory and in the sector $\left(
\partial ^{\mu },\partial ^{4}\right) $ it coincides with the operator $%
\overline{D}$ defined in (\ref{B4}) for the noncommutative tangent space
algebra $\Re _{\ell ,\infty }$.

The search for stability of the physical models describes well the evolution
of our understanding of physics. The transition from singular to generic
stable structures seems to tells us how Nature organizes itself. From
classical to relativistic mechanics we come from an unstable Galilean
algebra to the stable Lorentz algebra. Quantum mechanics also, may be
interpreted as the stabilization of the phase-space Poisson to the Moyal
algebra or, equivalently to the Heisenberg algebra. Finally, stabilizing the
Poincar\'{e}-Heisenberg algebra requires both a fundamental length ($\ell $)
and the noncommutativity of the translations ($\phi $). There is a formal
identity of all these processes and the last one relates to the emergence of
gravity. Of course, this is only a sketch of my understanding of the deep
intuition of David Finkelstein when stating that "gravity is a quantum
effect".

\section{Appendix: Irreducible representations of the space-time algebra}

For $SO\left( 3,2\right) $ $\left( \epsilon _{4}=+1\right) $\ a way to
characterize the irreducible representations of this groups is to consider
its action on functions on a $V^{3,2}$ cone, with coordinates%
\begin{equation}
\begin{array}{lll}
y_{1} & = & e^{s}\cos \varphi _{2} \\ 
y_{2} & = & e^{s}\sin \varphi _{2}\cos \varphi _{1} \\ 
y_{3} & = & e^{s}\sin \varphi _{2}\sin \varphi _{1} \\ 
y_{4} & = & e^{s}\sin \theta _{1} \\ 
y_{0} & = & e^{s}\cos \theta _{1}%
\end{array}
\label{B6}
\end{equation}%
Then, on this cone, consider a space $S^{\sigma ,\varepsilon }$ of functions
satisfying the homogeneity conditions \cite{Vilen}%
\begin{equation}
f\left( ax\right) =\left\vert a\right\vert ^{\sigma }sign^{\varepsilon
}af\left( x\right)  \label{B8}
\end{equation}%
$\sigma \in \mathbb{R}$ and $\varepsilon =\left\{ 0,1\right\} $. In $%
M^{\sigma ,\varepsilon }$ the group operators act as follows%
\begin{equation}
T\left( g\right) f\left( x\right) =f\left( g^{-1}x\right)  \label{B9}
\end{equation}%
Because of (\ref{B8}) the functions are uniquely characterized by their
values in the $\left( s=0\right) $ $\Gamma _{1}$ contour. This contour is
topologically $S^{2}\times S^{1}$. The spaces of homogeneous functions on
this contour will be denoted $S^{\Gamma _{1}}$. Denote by $g_{ij}\left(
\theta \right) $ a rotation in the plane $ij$ and by $g_{ij}^{\prime }\left(
t\right) $ a hyperbolic rotations in the plane $ij$.

Given $f\in S^{\Gamma _{1}}$, using (\ref{B8}) and (\ref{B9}) one obtains
for an hyperbolic rotation in the $1,4$ plane%
\begin{eqnarray}
T^{\sigma }\left( g_{14}^{\prime }\left( t\right) \right) f\left( \varphi
_{1},\varphi _{2},\theta _{1}\right) &=&\left\vert a\right\vert ^{\sigma
/2}f\left( \varphi _{1},\varphi _{2}^{\prime },\theta _{1}^{\prime }\right) 
\notag \\
\left\vert a\right\vert &=&\left\{ \sin ^{2}\theta _{1}+\left( \cos \theta
_{1}\cosh t-\cos \varphi _{2}\sinh t\right) ^{2}\right\} ^{1/2}  \notag \\
\cos \varphi _{2}^{\prime } &=&\frac{\cos \varphi _{2}\cosh t-\cos \theta
_{1}\sinh t}{\left\vert a\right\vert }  \notag \\
\cos \theta _{1}^{\prime } &=&\frac{\cos \theta _{1}\cosh t-\cos \varphi
_{2}\sinh t}{\left\vert a\right\vert }  \label{B10}
\end{eqnarray}%
Similar expressions are obtained for the other elementary operations. From
these one obtains, as infinitesimal generators, a representation for the
generators of the algebra $\left\{ X_{\mu },M_{\mu \nu }\right\} $ as
operators in $S^{\Gamma _{1}}$%
\begin{eqnarray}
iX_{1} &=&iM_{14}=\sigma \cos \theta _{1}\cos \varphi _{2}-\sin \theta
_{1}\cos \varphi _{2}\frac{\partial }{\partial \theta _{1}}-\cos \theta
_{1}\sin \varphi _{2}\frac{\partial }{\partial \varphi _{2}}  \notag \\
iX_{2} &=&iM_{24}=\sigma \cos \theta _{1}\sin \varphi _{2}\cos \varphi
_{1}-\sin \theta _{1}\sin \varphi _{2}\cos \varphi _{1}\frac{\partial }{%
\partial \theta _{1}}  \notag \\
&&+\cos \theta _{1}\cos \varphi _{2}\cos \varphi _{1}\frac{\partial }{%
\partial \varphi _{2}}-\frac{\cos \theta _{1}\sin \varphi _{1}}{\sin \varphi
_{2}}\frac{\partial }{\partial \varphi _{1}}  \notag \\
iX_{3} &=&iM_{34}=\sigma \cos \theta _{1}\sin \varphi _{2}\sin \varphi
_{1}-\sin \theta _{1}\sin \varphi _{2}\sin \varphi _{1}\frac{\partial }{%
\partial \theta _{1}}  \notag \\
&&+\cos \theta _{1}\cos \varphi _{2}\sin \varphi _{1}\frac{\partial }{%
\partial \varphi _{2}}+\frac{\cos \theta _{1}\cos \varphi _{1}}{\sin \varphi
_{2}}\frac{\partial }{\partial \varphi _{1}}  \notag \\
iX_{0} &=&iM_{04}=\frac{\partial }{\partial \theta _{1}}  \notag \\
iM_{12} &=&-\cos \varphi _{1}\frac{\partial }{\partial \varphi _{2}}+\frac{%
\cos \varphi _{2}\sin \varphi _{1}}{\sin \varphi _{2}}\frac{\partial }{%
\partial \varphi _{1}}  \notag \\
iM_{13} &=&-\sin \varphi _{1}\frac{\partial }{\partial \varphi _{2}}-\frac{%
\cos \varphi _{2}\cos \varphi _{1}}{\sin \varphi _{2}}\frac{\partial }{%
\partial \varphi _{1}}  \notag \\
iM_{23} &=&-\frac{\partial }{\partial \varphi _{1}}  \notag \\
iM_{10} &=&\sigma \sin \theta _{1}\cos \varphi _{2}+\cos \theta _{1}\cos
\varphi _{2}\frac{\partial }{\partial \theta _{1}}-\sin \theta _{1}\sin
\varphi _{2}\frac{\partial }{\partial \varphi _{2}}  \notag \\
iM_{20} &=&\sigma \sin \theta _{1}\sin \varphi _{2}\cos \varphi _{1}+\cos
\theta _{1}\sin \varphi _{2}\cos \varphi _{1}\frac{\partial }{\partial
\theta _{1}}  \notag \\
&&+\sin \theta _{1}\cos \varphi _{2}\cos \varphi _{1}\frac{\partial }{%
\partial \varphi _{2}}+\frac{\sin \theta _{1}\sin \varphi _{1}}{\sin \varphi
_{2}}\frac{\partial }{\partial \varphi _{1}}  \notag \\
iM_{30} &=&\sigma \sin \theta _{1}\sin \varphi _{2}\sin \varphi _{1}+\cos
\theta _{1}\sin \varphi _{2}\sin \varphi _{1}\frac{\partial }{\partial
\theta _{1}}  \notag \\
&&+\sin \theta _{1}\cos \varphi _{2}\sin \varphi _{1}\frac{\partial }{%
\partial \varphi _{2}}+\frac{\sin \theta _{1}\cos \varphi _{1}}{\sin \varphi
_{2}}\frac{\partial }{\partial \varphi _{1}}  \label{B11}
\end{eqnarray}

These representations are irreducible for non-integer $\sigma $. There are
also conditions for unitary of the representations, but this is not so
important because only the $M_{\mu \nu }$ $\left( \mu ,\nu =0,1,2,3\right) $
are generators of symmetry operations.

A similar construction is possible for $SO\left( 4,1\right) $ $\left(
\epsilon _{4}=-1\right) $\ with functions on a $V^{4,1}\left( \epsilon
_{5}=-1\right) $ cone, with coordinates%
\begin{equation}
\begin{array}{lll}
y_{1} & = & e^{s}\cos \varphi _{3} \\ 
y_{2} & = & e^{s}\sin \varphi _{3}\cos \varphi _{2} \\ 
y_{3} & = & e^{s}\sin \varphi _{3}\sin \varphi _{2}\cos \varphi _{1} \\ 
y_{4} & = & e^{s}\sin \varphi _{3}\sin \varphi _{2}\sin \varphi _{1} \\ 
y_{0} & = & e^{s}%
\end{array}
\label{B12}
\end{equation}%
the contour $\Gamma _{2}$ $\left( s=0\right) $\ in this case being
topologically $S^{3}$.

\end{document}